\newif\ifdraft\drafttrue
\newif\ifcolor\colortrue
\documentclass{usiinftr}
\pdfoutput=1 
\newcommand{\smartparagraph}[1]{\noindent{\bf #1}\ }
\newcommand{\eg}{{\it e.g.}}
\newcommand{\ie}{{\it i.e.}}

\usepackage{xcolor}
\definecolor{pennblue}{cmyk}{1,.65,0,.3}
\definecolor{pennred}{cmyk}{0,1,.65,.34}

\graphicspath{{./}}

\usepackage{cmtt}

\usepackage{paralist}
\usepackage{lastpage}
\usepackage{times}
\usepackage{tikz}
\usepackage{alltt}
\usepackage{graphicx}
\usepackage{balance}
\usepackage{amsmath}
\usepackage{caption}
\usepackage{subcaption}
\usepackage{algorithm,algpseudocode}
\usepackage{xspace}

\usepackage{listings}

\begin{document}

\title{Paxos Made Switch-y}

\author{Huynh Tu Dang}{1}
\author{Marco Canini}{2}
\author{Fernando Pedone}{1}
\author{Robert Soul\'{e}}{1}

\affiliation{1}{\USIINF}
\affiliation{2}{Universit\'{e} catholique de Louvain,  Belgium}

\TRnumber{2015-05}

\date{}

\maketitle

\begin{abstract}
This paper describes an implementation of the well-known consensus protocol, Paxos, in the P4 programming language. P4 is a language
for programming the behavior of network forwarding devices (i.e., the network data plane). Moving consensus logic into network devices could significantly improve the performance of the core infrastructure and services in data centers. Moreover, implementing Paxos in P4 provides a critical use case and set of requirements for data plane language designers.  In the long term, we imagine that consensus could someday be offered as a network service, just as point-to-point communication is provided today. 

\end{abstract}

\section{Introduction}
\label{sec:introduction}

Paxos~\cite{lamport98} is one of the most widely used protocols for solving
\emph{consensus}, the problem of getting a group of participants to reliably
agree on some value used for computation. Paxos is used to implement state
machine replication~\cite{lamport78,schneider90}, which is the foundation for
building fault-tolerant systems, including many of the core infrastructure
systems and services deployed in data centers, such as
OpenReplica~\cite{openreplica}, Ceph~\cite{ceph}, and Google's
Chubby~\cite{burrows06}.  Since most data center applications critically depend
on these services, the performance of Paxos has a dramatic impact on the overall
performance of the data center.

In prior work~\cite{dhang15}, we argued that significant performance
improvements could be gained by moving Paxos logic into network forwarding
devices.  Specifically, offering consensus as a network service would both
reduce the number of hops that consensus messages need to travel, and remove
message-processing bottlenecks at servers. As part of that work, we identified a
sufficient set of operations that a switch would need to perform in order to
implement Paxos logic. However, until recently, implementing Paxos logic inside
of a network switch would be challenging, potentially requiring coordination
with a particular vendor, and a customized hardware implementation. At the
present time, the landscape for network computing hardware has begun to change.
Several devices are on the horizon that offer flexible hardware with
customizable packet processing pipelines, including PISA chips from Barefoot
networks~\cite{bosshart13} and FlexPipes from Intel~\cite{jose15}. Other vendors
such as Cisco and Cavium will soon produce similar devices.

This new hardware presents exciting opportunities for language designers, as
they explore the question: \emph{what are the right abstractions for programming
forwarding devices, and how should those abstractions compose?}
A handful of recent projects have made initial proposals, including Huawei's
POF~\cite{song13}, Xilinx's PX~\cite{brebner14}, and the P4 Consortium's
P4~\cite{bosshart14}. These languages are poised to significantly improve the
flexibility and programmability of the network data plane. However, to
demonstrate their practical value, more work is needed around new applications
and use cases.

In this paper, we describe an implementation of Paxos in the P4
language~\cite{bosshart14}. Our choice for P4 is pragmatic: the language is open
and relatively more mature than other alternatives. Although Paxos is a
relatively simple protocol, there are many details that make an implementation
challenging. Consequently, there has been a rich history of research papers that
describe Paxos implementations, including attempts to make Paxos Simple
~\cite{lamport01}, Practical~\cite{mazieres07}, Moderately
Complex~\cite{vanrenesse15}, and Live~\cite{chandra07}. This paper differs from
the prior literature because implementing Paxos on packet forwarding devices
introduces new practical concerns that have not, to the best of our knowledge,
been previously addressed.

Overall, we make the following three contributions: First, we expose a new area
of research for consensus protocol designers, made interesting by the
restrictions of the target platform (e.g., constraints on field size, persistent
storage history, etc.). Second, we present an interesting use case for network
hardware programming languages, as the Paxos protocol requires packet processing
with a relatively complex logic that goes far beyond the examples published in
existing literature~\cite{bosshart14}. Third, to users of P4, our experience
with implementing Paxos provides a useful, concrete example of techniques that
can be applied to other data plane applications.

All source code, as well as a demo running in Mininet, is publicly available
under an open source license\footnote{https://github.com/usi-systems/p4paxos}.

The rest of this paper is organized as follows. We first provide short summaries
of the Paxos protocol (\S\ref{sec:background}) an the P4 language
(\S\ref{sec:p4}). We then discuss our implementation in detail
(\S\ref{sec:implementation}), followed by a general discussion of optimizations,
challenges, and future work (\S\ref{sec:discussion}).  Finally, we discuss
related work (\S\ref{sec:related}), and conclude (\S\ref{sec:conclusion}).

\section{Paxos Background}
\label{sec:background}

\begin{figure}
\centering
 \includegraphics[width=0.4\columnwidth]{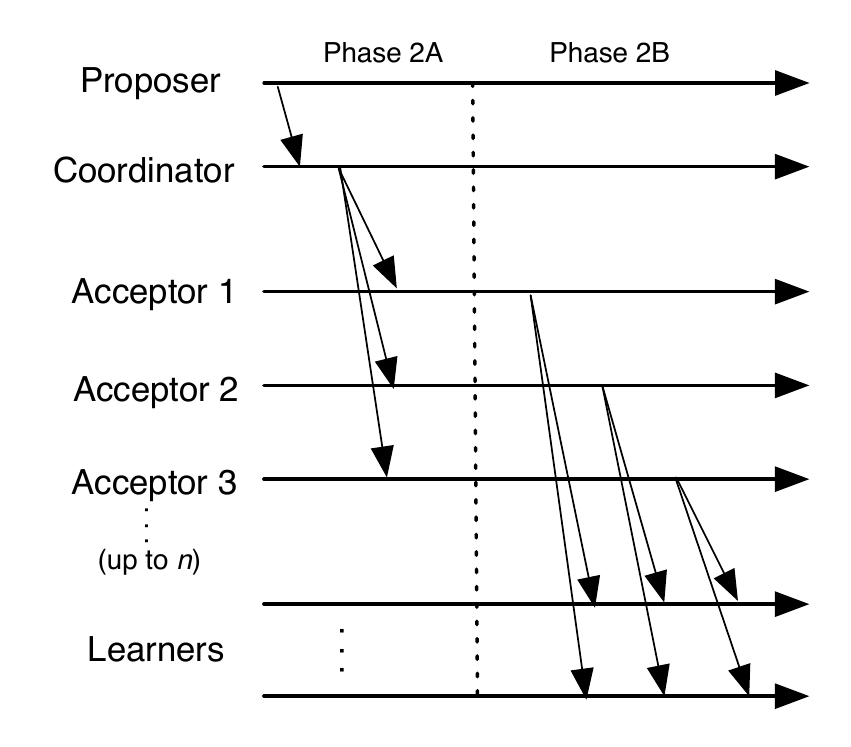}
\caption{The Paxos protocol communication pattern.}
\label{fig:paxos}
\end{figure}

State-machine replication is used to replicate services, so that a
failure at any one replica does not prevent the remaining operational replicas from
servicing client requests. State-machine replication is implemented using a
\emph{consensus} protocol, which dictates how the participants propagate and
execute commands.

Paxos~\cite{lamport98} is perhaps the most widely used consensus protocol. The
participants are processes that communicate by exchanging messages. Processes
may simultaneously play one or more of fours roles: \emph{proposers} issue
requests to the distributed system (\ie, propose a value); \emph{coordinators}
establish an ordering of requests; \emph{acceptors} choose a single value; and
\emph{learners} provide replication by learning what value has been chosen.

A Paxos \emph{instance} is one execution of consensus. An instance begins when a
proposer issues a request, and ends when learners know what value has been
chosen by the acceptor.  The protocol proceeds in a sequence of rounds. Each
round has two phases.  For each round, one process, typically a proposer or
acceptor, acts as the \emph{coordinator} of the round. The coordinator is
selected via an application-specific protocol, called \emph{leader election},
which is external to the Paxos protocol.


\smartparagraph{Phase 1.}
The coordinator selects a unique round number \textit{rnd} and asks the
acceptors to promise that in the given instance they will reject any requests
(Phase 1 or 2) with round number less than \textit{rnd}. Phase 1 is completed
when a majority-quorum $Q_a$ of acceptors confirms the promise to the
coordinator. If any acceptor already
accepted a value for the current instance, it will return this value to the
coordinator, together with the round number received when the value was accepted
(\textit{vrnd}).

\smartparagraph{Phase 2.} 
Figure~\ref{fig:paxos} illustrates the communication pattern of the Paxos
participants during Phase 2.
  The coordinator selects a value according to the
following rule: if no acceptor in $Q_a$ accepted a value, the coordinator can
select any value. If however any of the acceptors returned a value in Phase 1,
the coordinator is forced to execute Phase 2 with the value that has the highest
round number \textit{vrnd} associated to it. In Phase 2, the coordinator sends
a message containing a round number (the same used in Phase 1). Upon receiving
such a request, the acceptors acknowledge it, unless they have already
acknowledged another message (Phase 1 or 2) with a higher round
number. Acceptors update their \textit{rnd} and \textit{vrnd} variables with
the round number in the message. When a quorum of acceptors accepts the same
round number (Phase 2 acknowledgment), consensus terminates: the value is
permanently bound to the instance, and nothing will change this decision. Thus,
learners can deliver the value. Learners learn this decision either by
monitoring the acceptors or by receiving a decision message from the
coordinator.

\vspace*{3mm}
As long as a nonfaulty coordinator is eventually selected and there is a majority
quorum of nonfaulty acceptors and at least one nonfaulty proposer, every
consensus instance will eventually decide on a value. A failed coordinator is
detected by the other nodes, which select a new coordinator. If the coordinator does not receive a
response to its Phase 1 message it can re-send it, possibly with a bigger round
number. The same is true for Phase 2, although if the coordinator wants to execute
Phase 2 with a higher round number, it has to complete Phase 1 with that round
number.

The above describes one instance of Paxos. Throughout this
paper, references to Paxos implicitly refer to multiple instances chained together
(\ie, Multi-Paxos~\cite{chandra07}).

\section{P4 Background}
\label{sec:p4}

P4~\cite{bosshart14} is a network data plane programming language. Its design is
motivated both by the evolving OpenFlow standard~\cite{mckeown08}, and by the
need of many data centers for customized data plane functionality, for example,
to simplify network management or enable data-center specific features. In
contrast to other hardware programming languages, such as Verilog, P4 provides
higher-level abstractions that are tailored directly to the needs of network
forwarding devices. A complete language specification is available online~\cite{p4spec}.

The P4 program describes a sequence of tables, which \emph{match} on packet
header fields, and perform \emph{actions} that forward, drop, or modify packets.
The P4 language presents developers with five core abstractions:

\begin{enumerate}
\item \emph{Header Fields:} A packet header is a collection of
 fields. Developers must specify the width of each field. At most one field may
 be variable length.
\item \emph{Parsers:} Parsers describe how to transform packets to a
  parsed representation, from which header instances may be extracted.
\item \emph{Tables:} Developers specify which fields are examined from each
  packet, how those fields are matched, and actions performed as a consequence
  of the matching.
\item \emph{Actions:} Actions are invoked by tables, which are used to modify
  fields; add or remove headers; drop or forward packets; or perform stateful memory operations.
\item \emph{Control:} Developers specify how tables are composed.
\end{enumerate}

Beyond these five basic abstractions, P4 offers additional language constructs
for performing stateful operations. Our implementation of Paxos uses
\emph{registers} and \emph{metadata}. Registers provide persistent state
organized into an array of \emph{cells}. When declaring a register, developers
specify the size of each cell, and the number of cells in the array.  Metadata
is used in a similar way to registers, but provides a mechanism for storing
volatile per-packet state that may not be derivable from the packet header.

\section{P4 Paxos}
\label{sec:implementation}

\begin{figure}
\centering
 \includegraphics[width=0.4\columnwidth]{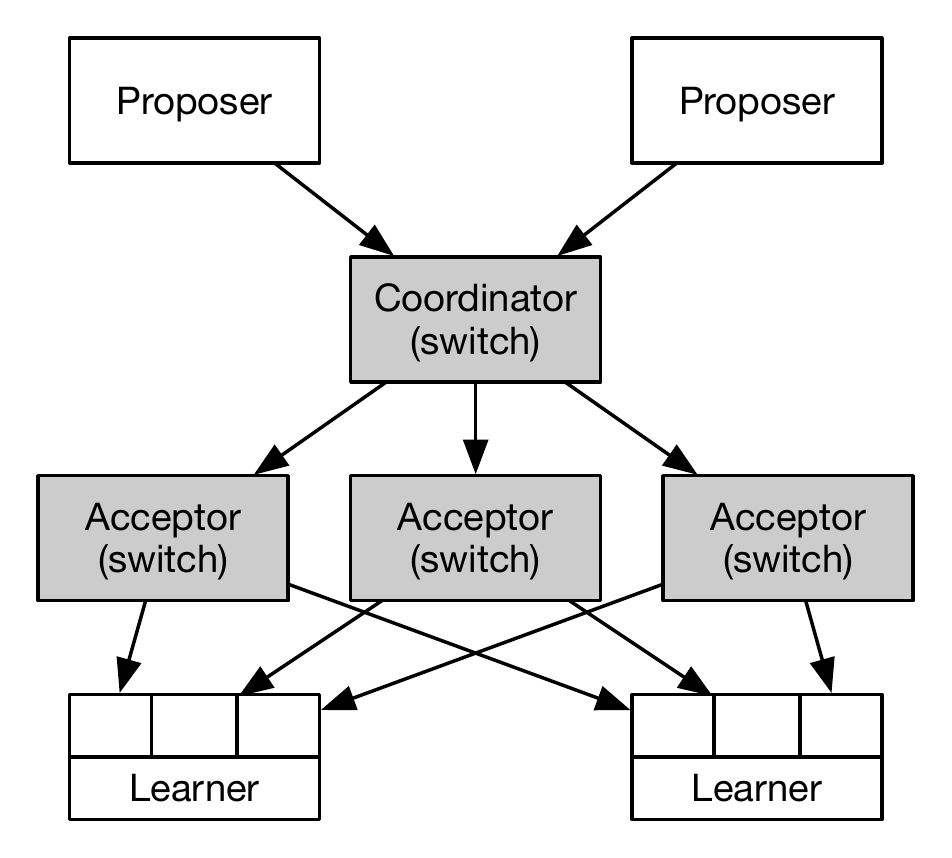}
\caption{A switch-based Paxos architecture. Switch hardware is shaded grey. Other
  devices are commodity servers. The learners each have four network interface cards.}
\label{fig:netpaxos}
\end{figure}  

Figure~\ref{fig:netpaxos} illustrates the architecture of switch Paxos. Switch
hardware is shaded grey, and commodity servers are colored white. As with
traditional Paxos, there are four roles that participants in the protocol play:
proposers, coordinators, acceptors, and learners. In switch Paxos, the
functionality of \emph{coordinators} and \emph{acceptors} is executed on
forwarding devices.

Although the Paxos protocol described in Section~\ref{sec:background}
 describes two phases, Phase 1 does not depend on any particular
 value, and can be run in advance for a large bounded number of
 values~\cite{lamport98}. The pre-computation needs to be re-run under two scenarios:
 either $(i)$ the Paxos instance approaches the bounded number of
 values, or $(ii)$ the device acting as coordinator changes (possibly
 due to failure). Therefore, it is common for Paxos
 implementations to only implement Phase 2. Our switch-based Paxos follows this
 same approach.
 
The illustration in Figure~\ref{fig:netpaxos} only shows one coordinator.  If
the other participants in the Paxos protocol suspect that the switch is faulty,
the coordinator functionality can be moved to either another switch in the
network or a server that temporarily assumes the role of coordinator. The
specifics of the leader-election process are application-dependent. We have
elided these details from the figure in order to simplify the presentation of
our design.


\paragraph*{Paxos Header}
\label{sec:header}

All packets used for communication in switch Paxos must include a Paxos-specific
protocol header, which is encapsulated in a UDP packet. Figure~\ref{fig:header}
shows the P4 specification of the packet header and parser.
The header contains the following fields:

\begin{itemize}
\item \texttt{msgtype}: distinguishes the various Paxos messages (e.g., 1A, 2A, etc.).
\item \texttt{inst}: the consensus instance of the message.
\item \texttt{rnd}: the semantics depend on who is sending the message. If the
  proposer sends the message, it is the round number computed in Phase 1. If the
  acceptor sends the message, it is the round number the acceptor has cast a vote for a proposal.
\item \texttt{vrnd}: the round number in which an acceptor has cast a vote.
\item \texttt{value}: the proposed value, or the value for which an acceptor has cast a vote. 
\end{itemize}

Note that message types are implemented with \texttt{\#define} macros
since there is no notion of an enumerated type in P4.

There are a few interesting decisions in the design of this header. First, the
number of instances that can be pre-computed in Phase 1 is bound by the size of
the \texttt{inst} field. If this field is too small, then consensus could only
be run for a short time. On the other hand, the coordinator and acceptor code
must reserve sufficient memory and make comparisons on this value, so setting the field too big could
potentially impact performance
Second, ideally, the \texttt{value} would be
stored in the packet payload, not the packet header. However, Paxos must
maintain the history of values, and to do so in P4, the field must be parseable,
and stored in a register. Therefore, our current implementation has
\texttt{value} in the header. Third, not all values in \texttt{value} will have
the same size. This size is dependent on the application. While P4 plans to
support variable length fields, the current implementation only supports fixed
length fields. Since we have to conservatively set the value to the size of the
largest value, we are storing potentially unused bytes.

\begin{figure}[t]
\centering
\begin{lstlisting}[xleftmargin=5.0ex]
header_type paxos_t {
    fields {
        msgtype : 8;  
        inst    : INST_SIZE;
        rnd     : 8;    
        vrnd    : 8;   
        value   : VALUE_SIZE;   
    }
}

parser parse_paxos {
    extract(paxos);
    return ingress;
}

\end{lstlisting}
\caption{Paxos packet header and parsers.}
\label{fig:header}
\end{figure}

\paragraph*{Proposer}
\label{sec:proposer}
The proposer logic is implemented as a library that exposes a simple API to the
application. The API consists of a single method \texttt{submit}, which is used
by the application to submit values and receive responses. The proposer
component receives requests from the application, and creates a switch Paxos
message to be sent to the coordinator (i.e., a Phase 2A message).  Messages are
sent to an IP Multicast address, which allows the coordinator and acceptors to
efficiently multicast messages to multiple destinations. We define the multicast group
address and associate the ports to it as part of an external switch
configuration. When packets are processed at switches, the P4 runtime handles
duplicating packets and forwarding out multiple ports.  Our prototype is
implemented as a Python module which runs on a commodity server.

\paragraph*{Coordinator}
\label{sec:coordinator}

In Paxos, a coordinator brokers requests on behalf of proposers. Their role is
to impose an ordering of messages when multiple proposers concurrently propose
values. When there is a single coordinator, as is the case in our prototype, a
monotonically increasing sequence number can be used to order the messages. This
sequence number is written to the message's \texttt{inst} field. 

Thus, to implement coordinator logic in P4, the code needs to perform the
following actions: (i) copy the next-in-use instance number into the message
header, (ii) increment the instance number, and (iii) store the value of the new
instance number.

Figure~\ref{fig:coordinator} shows the corresponding P4 implementation. To
persist the value of the instance number, we use a register named
\texttt{reg\_inst} (lines 1-3). Recall that in P4, actions can only be initiated
in response to \emph{matches} in a table. Therefore, when a coordinator receives
a valid Paxos message, it will pass the packet to the table (lines 20-21),
\texttt{tbl\_sequence}, which will match on Paxos Phase 2A message and
perform the \texttt{handle\_2a} action. The
\texttt{handle\_2a} action performs the following operations: First it
reads the instance number from the register and writes the number to the
\texttt{inst} header field (line 7).  Next, it increments the sequence number,
and writes the updated value to the register (lines 8-9).  Finally, the
\texttt{tbl\_sequence} table passes control of the packet back to the
\texttt{control} block, which forwards the packet out on the appropriate ports.

\begin{figure}
\centering
\begin{lstlisting}[xleftmargin=5.0ex]
register reg_inst {
    width : INST_SIZE;
    inst_count : 1;
}

action handle_2a() {
    register_read(paxos.inst, reg_inst, 0);
    add_to_field(paxos.inst, 1);
    register_write(reg_inst, 0, paxos.inst);
}

table tbl_sequence {
    reads { paxos.msgtype : exact; }
    actions { handle_2a; _nop; }
    size : 1;
}

control ingress {
    /* process other headers */                                 
    if (valid(paxos)) {         
        apply(tbl_sequence);    
    }
}

\end{lstlisting}
\caption{Coordinator code.}
\label{fig:coordinator}
\end{figure}

\paragraph*{Acceptor}
\label{sec:acceptor}

Paxos acceptors receive messages from the coordinator, and decide whether to
accept or reject a proposal. Thus, acceptors are vital to the protocol for
ensuring the consistency of the whole system. To perform their functionality,
acceptors must maintain and access the history of proposals that they have
accepted. This state does not need to grow unbounded, though, as it may be
periodically trimmed. We have not included this ``cleanup'' in our P4
implementation.

When an acceptor receives a message, it must read the latest round number for
the current instance from its storage, and compare its round number to the round
number in the arriving packet. If the message is for a larger round number than
what observed so far, the acceptor must process is according to the message
type: either Phase 1A or Phase 2A. If it receives a Phase 1A message, it must
update its local round register with the contents of the arriving packet. When
an acceptor receives a Phase 2A message, it must update its state and forward
the message.

Figure~\ref{fig:acceptor} shows the P4 implementation of acceptor logic. The
acceptor needs to maintain several stateful constructs. It uses the metadata
\texttt{meta\_paxos} to store the round number for which it has cast a vote
(line 7), and three registers, indexed by consensus instance, to store the
history of rounds, vrounds, and values (lines 9-21).

The entry point to the acceptor logic is the control block (line 49). When a
packet arrives, it is passed to two tables.  The first table, \texttt{tbl\_rnd}
(line 42), invokes the \texttt{read\_rnd} action to copy the current round for
the instance specified in the packet to the metadata (lines 24-26). Then, the
stored round is compared to the round in the packet header (line 53). If the
round in the packet header is greater than the stored round, the packet is
passed to the acceptor table (line 54). Otherwise, the packet is dropped (line
55).

The acceptor table invokes two possible actions, corresponding to Paxos message
types.  The action, \texttt{handle\_1a}, sets the message type to Phase 1B,
reads the vrnd and the value from the registers, writes them on the
corresponding fields on the packet, and updates the round register with the
round it has just seen (lines 28-33). The action, \texttt{handle\_2a}, accepts
the proposal by updating the \texttt{rnds}, \texttt{vrnds} and \texttt{values}
registers based on the corresponding fields on the packet header, and updates
the message type from Phase 2A to Phase 2B.

\begin{figure}[t!]
\begin{lstlisting}[xleftmargin=5.0ex]
header_type paxos_metadata_t {
    fields {
        rnd : 8;
    }
}

metadata paxos_metadata_t meta_paxos;

register rnds {
    width      : 8;
    inst_count : NUM_INST;
}

register vrnds {
    width      : 8;
    inst_count : NUM_INST;
}

register values {
    width      : VALUE_SIZE;
    inst_count : NUM_INST;
}

action read_rnd() {
    register_read(meta_paxos.rnd, rnds, paxos.inst); 
}

action handle_1a() {
    modify_field(paxos.msgtype, PAXOS_1B);                                      
    register_read(paxos.vrnd, vrnds, paxos.inst);
    register_read(paxos.value, values, paxos.inst);  
    register_write(rnds, paxos.inst, paxos.rnd);                 
}

action handle_2a() {
    modify_field(paxos.msgtype, PAXOS_2B);				        
    register_write(rnds, paxos.inst, paxos.rnd);               
    register_write(vrnds, paxos.inst, paxos.rnd);              
    register_write(values, paxos.inst, paxos.value);               
}

table tbl_rnd { actions { read_rnd; } }

table tbl_acceptor {
    reads   { paxos.msgtype : exact; }
    actions { handle_1a; handle_2a; _drop; }
}

control ingress {
    /* process other headers */
    if (valid(paxos)) {          
        apply(tbl_rnd);
        if (meta_paxos.rnd <= paxos.rnd) {
            apply(tbl_acceptor);
         } else apply(tbl_drop);
     }
}
\end{lstlisting}
\caption{Acceptor code.}
\label{fig:acceptor}
\end{figure}

\paragraph*{Learner}
\label{sec:learner}

Learners are used by the protocol to provide replication by learning the result
of a consensus instance. Learners must receive votes from a majority-quorum of
the acceptors. This could be achieved in various ways. In
Figure~\ref{fig:paxos}, and in our prototype implementation, learners are
directly connected to each acceptor on a different network interface. In an
alternative implementation, acceptors could add an ``acceptor id'' to the packet
header, and an additional switch could be used to demultiplex messages from the
acceptor switches.

We briefly describe our prototype implementation in more detail. Each learner is
connected to all acceptors via a separate network interface. The learner
identifies different acceptors by distinguishing which interface a packet
arrives on. When a Phase 2B message arrives, each learner extracts the instance
number, round, and value. The learner maintains a two-dimensional array to store
the messages, where the first index represents an instance, and the second
index represents an interface. In other words, if a message with instance number
$i$ arrives on interface $j$, it is stored as the $i$,$j$ element of the matrix.
For a given instance $i$, if the learner receives identical phase 2B messages
from a majority-quorum of acceptors, then the corresponding value $v$ is
decided.  A majority is equal to $f + 1$ where $f$ is the number of faulty
acceptors that can be tolerated.  After reaching the decision, the learner
executes the request and responds to the proposers.

\paragraph*{Optimizations}

Implementing Paxos in P4 requires $2f + 1$ acceptors. Considering that acceptors
in our design are network switches, this could be too demanding. However, we
note that one could exploit existing Paxos optimizations to spare resources.
Cheap Paxos~\cite{cheap-paxos} builds on the fact that only a majority-quorum of
acceptors is needed for progress.  Thus, the set of acceptors can be divided
into two classes: first-class acceptors, which would be implemented in the
switches, and second-class acceptors, which would be deployed in commodity
servers.  In order to guarantee fast execution, we would require $f + 1$
first-class acceptors (i.e., a quorum) and $f$ second-class acceptors.
Second-class acceptors would likely fall behind, but would be useful in case a
first-class acceptor fails. Another well-known optimization is to co-locate the
coordinator with an acceptor, which in our case would be an acceptor in the
first class.  In this case, a system configured to tolerate one failure ($f=1$)
would require only two switches.

\section{Discussion}
\label{sec:discussion}

The code in Section~\ref{sec:implementation} provides a relatively complex instance
of a data plane application that we hope can be useful to other P4 programmers.
However, beyond providing a concrete example, the process of implementing Paxos
in P4 also draws attention to requirements for P4 specifically, and data plane
languages in general. It also highlights future areas of research for designers
of consensus protocols. We expand the discussion of these two topics below. 

\subsection{Impact on P4 Language}

Implementing Paxos in P4 provides an interesting use case for data plane
programming languages. As a result of this experience, we developed several
``big-picture'' observations about the language and future directions for
extensions or research. We share the observations below, because it is valuable
to crystallize them in writing.

\paragraph*{Programming with tables}
P4 presents a paradigm of ``programming with tables'' to developers. This
paradigm is somewhat unnatural to imperative (or functional) programmers, and it
takes some time to get accustomed to the abstraction. It also, occasionally,
leads to awkward ways of expressing functionality. For example, the logic in the
acceptor code requires first reading the round number from storage, and then
perform a different action depending on the result of a comparison to the
current packet's round number. However, the round number cannot be read from
storage directly in the control statement. Instead, it must be performed as an
action that results from a table application (\texttt{tbl\_rnd}). It may be
convenient to allow storage accesses directly from \texttt{control} blocks.

\paragraph*{Modular code development}
Although P4 provides macros that allow source to be imported from other files
(e.g., \texttt{\#include}), the lack of a module system makes it difficult to
separate functionality, and build applications through composition, as is
usually suggested as best practice for software engineering. For example, it
would be nice to be able to ``import'' a Paxos module into an L2 learning
switch.  This need is especially acute in \texttt{control} blocks, where tables
and control flow have to be carefully arranged. As the number of tables, or
data plane applications, grows, it seems likely that developers will make
mistakes.

\paragraph*{Error handling}
Attempting to access a register value from an index that exceeds the size of the
array results in a segmentation fault. Obviously, performing bounds checks for
every memory access would add performance overhead to the processing of
packets. However, the alternative of exposing unsafe operations that could lead
to failures seems equally undesirable. It may be useful in the future to provide
an option to execute in a ``safe mode'', which would provide run-time boundary
checks as a basic precaution. It would also be useful to provide a way for
programs to catch and recover from errors or faults.

\paragraph*{Control of memory layout}
While P4 provides a stateful memory abstraction (a register), there is no
explicit way of controlling the memory layout across a collection of registers
and tables, and its implementation is target dependent. In our case, the
\texttt{tbl\_rnd} and \texttt{tbl\_acceptor} tables end up realizing a pipeline
that reads and writes the same shared registers. However, depending on the
target, the pipeline might be mapped by the compiler to separate memory or
processing areas that cannot communicate, implying that our application would
not be supported in practice. It would be helpful to have ``annotations'' to
give hints regarding tables and registers that should be co-located.

\paragraph*{Efficient hardware translation}
P4 is a quite powerful and expressive language. We were (pleasantly) surprised
at how easily we could express the relatively complex Paxos logic. However, we
have not yet been able to evaluate our Paxos implementation on an actual
hardware deployment, and it remains to be seen if new hardware can actually
support efficient implementations of P4 programs. Data plane language designers
face a challenge of balancing expressivity with performance. How to negotiate
this tradeoff remains an open question.



\subsection{Impact on Paxos Protocol}

Consensus protocols are typically designed without consideration for the
networks on which they run. As a result, most consensus protocols make weak
assumptions about network behavior, and therefore, incur overhead
to compensate for potential message loss or re-ordering. However, 
advances in network hardware programmability have laid a foundation for
designing new consensus protocols which leverage assumptions about network
computing power and behavior in order to optimize performance.

One potentially fruitful direction would be to take a cue from systems like Fast
Paxos~\cite{lamport06} and Speculative Paxos~\cite{ports15}, which take
advantage of ``spontaneous message ordering" to implement low-latency consensus.
Informally, spontaneous message order is the property that with high probability
messages sent to a set of destinations will reach these destinations in the same
order.  This can be achieved with a careful network configuration~\cite{ports15}
or in local-area networks when communication is implemented with
IP-multicast~\cite{PS02b}.


By moving part of the functionality of Paxos and its variations to switches,
protocol designers can explore different optimizations.  A switch could improve
the chances of spontaneous message ordering and thereby increase the likelihood
that Fast Paxos can reach consensus within few communication steps (i.e., low
latency).  Moreover, if switches can store and retrieve values, one could
envision an implementation of Disk Paxos~\cite{gafni00} that relies on stateful
switches, instead of storage devices.  This would require a redesign of Disk
Paxos since the storage space one can expect from a switch is much smaller than
traditional storage.

\section{Related Work}
\label{sec:related}

In prior work~\cite{dhang15}, we proposed the idea of moving consensus logic to
forwarding devices using two approaches: (i) implementing Paxos in switches, and
(ii) using a modified protocol, named \emph{NetPaxos}, which solves consensus
without switch-based computation by making assumptions about packet ordering. This
paper builds on that work by making the implementation of a switch-based Paxos
concrete. In the process, we identify areas for future research both for
 data plane programming languages and consensus protocol design.

\smartparagraph{Data plane programming languages.}
Several recent projects have proposed domain-specific languages for data plane
programming. Notable examples including Huawei's POF~\cite{song13}, Xilinx's
PX~\cite{brebner14}, and the P4~\cite{bosshart14} language used throughput this
paper. We chose to focus on P4 because (i) there is a growing community of
active users, and (ii) it is relatively more mature than the other
choices. However, the ideas for implementing Paxos in switches should generalize
to other languages.


\smartparagraph{Replication protocols.}
Research on replication protocols for high availability is quite mature.
Existing approaches for replication-transparent protocols, notably protocols
that implement some form of strong consistency (\eg, linearizability,
serializability) can be roughly divided into three classes \cite{CBPS10}: (a)
state-machine replication~\cite{lamport78,schneider90}, (b) primary-backup
replication~\cite{OL88}, and (c) deferred update replication~\cite{CBPS10}.

At the core of all classes of replication protocol discussed above, there lies a
message ordering mechanism. This is obvious in state-machine replication, where
commands must be delivered in the same order by all replicas, and in deferred
update replication, where state updates must be delivered in order by the
replicas.  In primary-backup replication, commands forwarded by the primary must
be received in order by the backups; besides, upon electing a new primary to
replace a failed one, backups must ensure that updates ``in-transit'' submitted
by the failed primary are not intertwined with updates submitted by the new
primary~(\eg, \cite{PF00}).

Although many mechanisms have been proposed in the literature to order messages
consistently in a distributed system~\cite{DSU04}, very few protocols have taken
advantage of network specifics.  Protocols that exploit \emph{spontaneous
message ordering} to improve performance are in this category (\eg,
\cite{lamport06,PS02b,PSUC02}).  The idea is to check whether messages reach their
destination in order, instead of assuming that order must be always constructed
by the protocol and incurring additional message steps to achieve it. We believe
that ordering protocols have much to gain (\eg, in performance, in simplicity)
by tightly integrating with the underlying network layer. Recent advances in
programmable network hardware make this research endeavor realizable.

\section{Conclusion}
\label{sec:conclusion}

The advent of flexible hardware and expressive data plane programming languages
will have a profound impact on networks. One possible use of this emerging
technology is to move logic traditionally associated with the application layer
into the network itself. In the case of Paxos, and similar consensus protocols,
this change could dramatically improve the performance of data center
infrastructure. In this paper, we have described an implementation of Paxos in
the P4 language. This implementation is a first step towards the continued
development and evolution of data plane language, that also opens the door for
new research challenges in the design of consensus protocols.

\noindent \textbf{Acknowledgments:} We thank Changhoon Kim for valuable feedback
on this work. This research is (in part) supported by European Union's Horizon
2020 research and innovation programme under the ENDEAVOUR project (grant
agreement 644960).

{\small
\bibliographystyle{abbrv}
\bibliography{main}
}

\end{document}